\documentclass[conference]{IEEEtran}

\newcommand{\ket}[1]{|#1\rangle} 
\newcommand{\op}[1]{|#1\rangle\langle#1|} 

\usepackage{array}
\newcolumntype{?}[1]{!{\vrule width #1}}
\usepackage{graphicx}
\usepackage{xcolor}

\usepackage{cite}
\usepackage{amsmath,amssymb,amsfonts}
\usepackage{amsthm}
\usepackage{algorithmic}

\usepackage{comment}
\usepackage{colortbl}
\usepackage{courier}
\newtheorem{proposition}{Proposition}
\newtheorem{theorem}{Theorem}
\AtBeginDocument{%
  }

\begin{document}

\title{Dynamic Resource Allocation with \\  Quantum Error Detection}

\author{\IEEEauthorblockN{Quinn Langfitt$^{1}$, Alvin Gonzales$^2$, Joshua Gao$^3$, Ji Liu$^2$, Zain H. Saleem$^2$, Nikos Hardavellas$^1$, Kaitlin N. Smith$^{1}$}
\IEEEauthorblockA{
\textit{$^1$Department of Computer Science, Northwestern University, Evanston, IL, USA}\\
\textit{$^2$Mathematics and Computer Science Division, Argonne National Laboratory, Lemont, IL, USA} \\ 
\textit{$^3$Department of Computer Science, Virginia Tech, Blacksburg, VA, USA}
 \\
}
}

\maketitle

\begin{abstract}
Quantum processing units (QPUs) are highly heterogeneous in terms of physical qubit performance. To add even more complexity, drift in quantum noise landscapes has been well-documented. This makes resource allocation a challenging problem whenever a quantum program must be mapped to hardware. As a solution, we propose a novel resource allocation framework that applies Pauli checks. Pauli checks have demonstrated their efficacy at error mitigation in prior work, and in this paper, we highlight their potential to infer the noise characteristics of a quantum system. Circuits with embedded Pauli checks can be executed on different regions of qubits, and the syndrome data created by error-detecting Pauli checks can be leveraged to guide quantum program outcomes toward regions that produce higher-fidelity final distributions. Using noisy simulation and a real QPU testbed, we show that dynamic quantum resource allocation with Pauli checks can outperform state-of-art mapping techniques, such as those that are noise-aware. Further, when applied toward the Quantum Approximate Optimization Algorithm, techniques guided by Pauli checks demonstrate the ability to increase circuit fidelity $>11\%$ on average, and up to $33\%$. 
\end{abstract}

\section{Introduction }

\begin{figure}[t]
     \centering
         \includegraphics[width=0.99\linewidth,trim={0cm 0cm 0cm 0cm},clip]{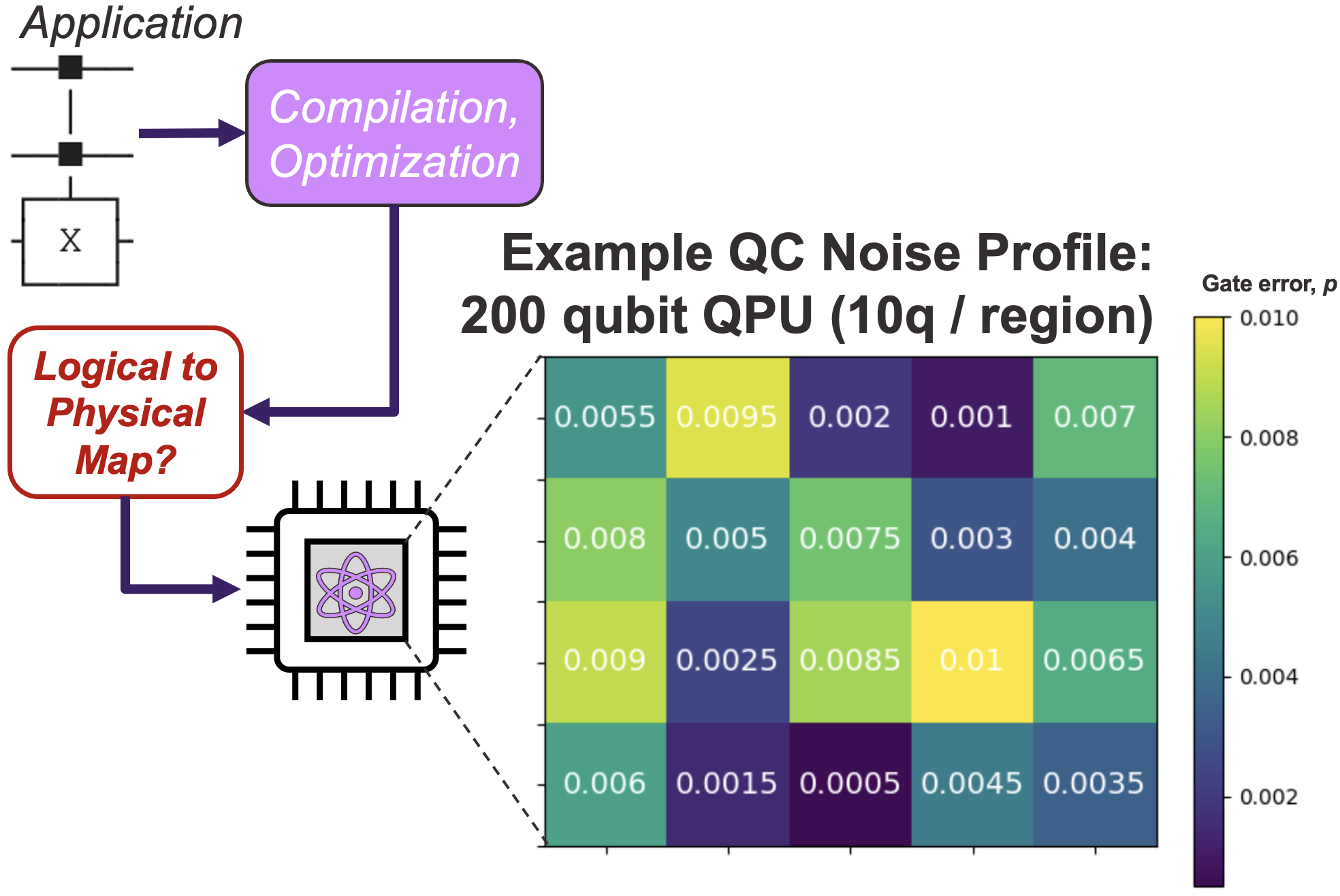}
        \caption{A challenging problem facing quantum programmers involves mapping applications to QPUs with heterogeneous noise profiles. Pictured here is a QPU with 20 regions (10 qubits per region) of unique 1q error rates, $p\in[0.0005,0.01]$, and 2q errors, $2p$. Our proposal leverages Pauli check based characterization to learns a QPU's noise when confidence in device properties are low.}
        \label{fig:noise-profile}
        
\end{figure}

Programming a quantum computer (QC) is a challenging task. First, arbitrary quantum programs must be compiled so that they respect QPU architectural constraints such as qubit-qubit connectivity and supported basis gate set. Then, the circuit must be mapped to a robust region of hardware, which can be a moving target on today's devices. The consequence of a poorly mapped circuit is a low fidelity result, which is sub-optimal with the rising cost to access quantum hardware through the quantum cloud~\cite{ravi2021quantum}. Thus, we are motivated to develop techniques that enable informed quantum resource allocation that work toward maximal utilization of hardware and improved fidelity of quantum program outcomes.
The problem of mapping an application to a highly diverse QPU noise profile is illustrated in Fig.~\ref{fig:noise-profile}.

Much work has been dedicated to improving both quantum program success and machine utilization during runtime in an effort to ease bottlenecks that hinder scaling. 
For example, noise-adaptive mapping is a state-of-art quantum circuit optimization technique used to place and route quantum programs on quantum hardware
~\cite{murali2019noise}. Noise-aware schemes, however, depend on 
accurate calibration data either from the QC provider or from high-overhead benchmarking.  
Even in the case that a snapshot of machine properties is available, QC noise is observed to fluctuate over time in ways that are challenging to capture with analytical models~\cite{dasgupta2021stability}. With this in mind, it is very unlikely to fully know a QCs error properties with high accuracy and confidence at any moment in time. Moreover, calibration data typically only reveal average error rates via randomized benchmarking \cite{ Knill_2008RandomBenchOfQG, Emerson_2005ScalNoiseEstWithRandUOp}, and the errors affecting a specific circuit can deviate significantly from the idealized noise model \cite{Proctor_2017WhatRandomBenchActuallyMeas, Helsen_2022GenFrameForRandomBenchmarking}. If used for hardware allocation, sub-optimal mappings can result.

The magnitude of average error rates prevents practical implementations of quantum error correction (QEC) on today's QPUs. Fortunately, quantum error mitigation (QEM) techniques, or techniques that require classical postprocessing, have shown to be extremely valuable for handling errors that occur on a quantum processor during runtime. Examples of QEM include zero noise extrapolation~\cite{giurgica2020digital}, Clifford data regression~\cite{czarnik2021error}, measurement error mitigation~\cite{nation2021scalable}, and Pauli checks~\cite{Debroy_2020ExtendedFlagGadgetsForLowOverCircVerif,gonzales2023quantum}. Pauli checks are of particular interest in this work as they produce a syndrome that when measured, provides information about whether or not an error was detected. This inspired the idea to leverage Pauli check syndrome data to learn about QPU noise, especially in the use case where calibration data is unavailable. 

In our work, we seek alternative methods to determine the noise profile of a QC when its properties are unknown. 
This work presents a novel framework that combines the power of circuit ensembling with Pauli checks for dynamic quantum resource allocation. In this work, we define resource allocation as the process of identifying hardware regions that are available for use, analyzing their performance, and using the performance insights to guide what resources should be prioritized for use. Syndrome data produced by the Pauli checks provides a method to roughly characterize QC error, and this data is leveraged to allow higher priority contributions to quantum program final results from the most robust regions on the QPU. 
To learn noise that is most important to the success of target applications, our techniques intelligently embed Pauli checks into the quantum applications of intrest. These programs are then executed across a QPU for noise profile characterization. 
Knowledge of the QPU's best regions creates a weighted ensemble result with non-trivial gain, found to be $> 11\%$ on average, and up to 33\%, during our experiments using the Quantum Approximate Optimization Algorithm (QAOA). 

Our contributions are as follows:
\begin{itemize}

    \item We provide a study of the benefit associated with Pauli checks for quantum error detection and mitigation, while highlighting the considerations for their practical implementation. Pauli check sandwiching (PCS) and ancilla-free Pauli checks (AFPC) are of focus. To the best of our knowledge, AFPC is evaluated here at the application level for the first time.
    \item We propose the use of Pauli checks for quantum device characterization. We develop a framework that combines ensembled circuits with Pauli check based characterization and error mitigation in a novel manner for improved runtime results. \item The concept of a Pauli check weighted distribution is introduced, along with its threshold and hybrid variations.
    \item We show that we either perform as well or outperform state-of-art noise-aware mapping techniques. Further, we are able to achieve high-fidelity program outcomes without the guidance of QPU-provided error profile.
    \item We demonstrate how dynamic quantum resource allocation with Pauli checks can be leveraged to significantly improve program success on a variety of quantum applications, including QAOA.
\end{itemize}

We speculate that the methods presented here will also be advantageous in the not-too-distant future -- as larger quantum processors are deployed on the quantum cloud, quantum resource providers may not provide fine-grained particulars about QC operational characteristics and error rates as we are familiar with today.

\section{Prior Work and Motivation}

QC hardware is noisy during use and expensive to access. As a result, much work has been dedicated to improving both quantum program success and machine utilization during runtime. This Section will discuss QC noise-adaptive mapping, multi-programming, and circuit-based training, three quantum program optimization techniques that push the frontier of existing hardware. We discuss pros and cons of these methods while presenting new questions that motivate the new contributions of this work. 

This Section does not provide an exhaustive introduction to quantum computation. We recommend that the reader refers to Ref.~\cite{nielsen2010quantum} for a more complete introduction to topics in quantum information processing and Ref.~\cite{ding2022quantum} for an overview of the fundamentals in quantum computing systems.

\subsection{Noise-adaptive Mapping}

There are a variety of different qubit technologies that exist today, such as superconducting circuits, trapped ions, and neutral atoms, among others. Each of these platforms have their respective benefits and costs, and they all demonstrate a wide range of performance characteristics in terms of fidelity, coherence, and gate times. Even QCs comprised of the same qubit technology tend to demonstrate performance differences. Due to a combination of imperfect fabrication procedures and intrinsic environmental conditions, each quantum system is characterized by a unique noise landscape during computation. This presents the challenge of determining how to best allocate logical qubits to physical qubits during runtime. 

Noise-adaptive mapping is a state-of-art mapping technique used to place and route quantum programs on quantum hardware~\cite{tannu2019not,murali2019noise, waring2024noiseawareutilityoptimization, Ji_2022CalibraAwareTranspForVQO}. Given a circuit and a target QC, tools like ~\texttt{mapomatic} optimize for the best low-noise device sub-graph using the QC's error rates. Noise-aware schemes, however, often depend on the availability of calibration data from the QC service provider. Further, these schemes also assume that the calibration data provided by the quantum hardware vendor is complete and correct. If trusted machine data is unavailable, protocols such as SABRE~\cite{li2019tackling} will be left without (or with misinformed) guidance for logical to physical qubit mapping, resulting in poor quantum program performance. Even in the case that machine characterization data is complete and derives from a reliable source, it is known that machine noise and performance characteristics have a tendency to fluctuate over time in ways that are challenging to capture with analytical models\cite{dasgupta2021stability}. If stale calibration data is used in this case, sub-optimal mappings result. 

Noise characterization has previously been used to improve the accuracy of an ensemble of hardware executed circuits. Quancorde \cite{ravi2022quancorde} utilizes a form of Clifford noise estimation circuits \cite{Urbanek_2021MitigDepolNOnQCwithNoiseEstCirc} to estimate the noise present in a circuit execution. The final distribution is then calculated as a weighted sum of the individual distributions.
In our work, we explore alternative methods to determine the noise profile of quantum hardware due to the challenges associated with depending on provider-supplied data. It is desirable to develop techniques with minimal overheads associated with gathering characterization data. In a work with similar motivations as our own, Ref.~\cite{ravi2022quancorde} attempts to characterize the noise that an application will experience on a target QPU through the use of easy-to-simulate Clifford-circuits that approximate the structure of the circuit of interest. However, this Clifford-based technique has drawbacks. First, additional QC runtime is required to collect characterization data from the Clifford circuits, and these characterization results must be compared to simulated Clifford circuit outcomes. Second, the actual payload circuits must be run soon after characterization data is retrieved to avoid sub-optimal results from machine property drift. Third, depending on target circuit structure, a Clifford approximation might result in a very different unitary that changes its noise sensitivity. These drawbacks inspire us to develop techniques that allow QC noise learning to be interleaved into the target circuits of interest.

\subsection{Maximizing QC Utilization}

Some QCs on today's quantum cloud contain more than 100 qubits, but low qubit coherence times and high operator errors prohibit the use of all of these qubits collectively for computation. As a result, QC applications are typically much smaller than the qubit capacity of the machine, leaving many QC qubits unused during runtime. Recent work has explored how to take advantage of these unused qubit resources to improve quantum program outcomes.

Ideally, compute resources are scheduled for maximum utilization. As the work in Ref.~\cite{wu2024reducing} highlights, parallelized quantum circuit execution, as compared to a sequential approach, offers the benefit of reduced runtime latency because a quantum circuit executed $n$ times simultaneously can reach a threshold number of observables in a distribution in $1/n$ time. However, work in Ref.~\cite{das2019case} highlights that multi-programming quantum hardware, especially machines existing today, is a non-trivial task as improper scheduling procedures can cause undesirable crosstalk between two different workloads. The observation of crosstalk is especially likely whenever two quantum programs of different duration are executed adjacently -- the measurement operations of the shorter circuit could impact the longer circuit that still has quantum circuit operations in-progress. If implemented with care, however, multi-programming can increase quantum system throughput without having a detrimental impact on program outcomes. 

We are interested in exploring the potential that multi- programming has for improving quantum program performance. The work in~\cite{tannu2019ensemble} demonstrates how ensembling techniques employed to create a distribution from many circuit mappings can help smooth errors, improving fidelity. Creating a cumulative distribution from many instances of the same application implemented in parallel could have a similar effect that boosts performance. 

\subsection{Pauli Check Sandwiching}

`Pauli Check Sandwiching' (PCS) is a technique used to detect and mitigate errors~\cite{gonzales2023quantum, Debroy_2020ExtendedFlagGadgetsForLowOverCircVerif}. Prior work in PCS has shown its viability for mitigating gate and
measurement errors in quantum circuits~\cite{qutracer} and for fault injection~\cite{pcs-fault-injection} that studies error in quantum circuits. In our work, we also find that PCS is a powerful tool for guiding QC characterization, and we make the novel contribution of combining PCS with ensembling. 

\begin{figure}[t]
     \centering
         \includegraphics[width=0.99\linewidth,trim={0cm .5cm .5cm .5cm},clip]{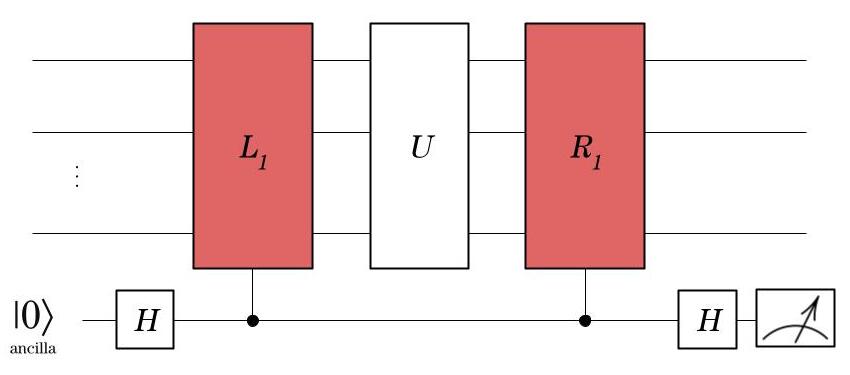}
        \caption{General PCS circuit layout. The red unitaries represent the Pauli checks that sandwich the main payload circuit. Measurement of the ancillas provide detection of errors that occurred in the payload circuit.}
        \label{fig:pcs-circuit}
        
\end{figure}

As seen in Fig.~\ref{fig:pcs-circuit}, PCS surrounds a payload circuit, $U$, with controlled Pauli operator checks. Errors on $U$ can be detected on an ancilla through phase kickback.
It is important that the relationship 

\begin{equation}\label{eq:pauli-check-facts}
    R_{1}UL_{1} = U
\end{equation}

\begin{figure}[t]
     \centering
         \includegraphics[width=0.99\linewidth,trim={0cm 0cm 0cm 0cm},clip]{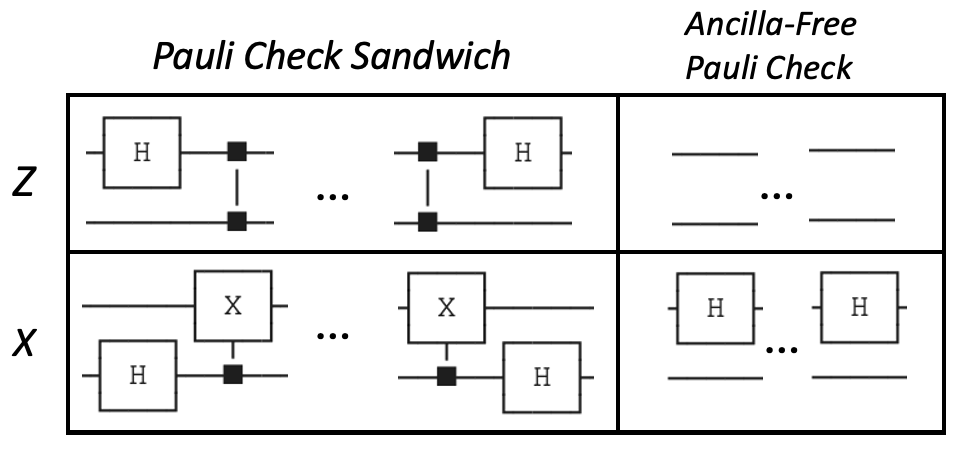}
        \caption{Example of the conversion between PCS and AFPC for $Z$ and $X$ checks. }
        \label{fig:pcs-to-afpc}
        
\end{figure}

\noindent holds so the introduced checks do not disturb the original payload circuit semantics. As a note, the operations contained in $L_1$ and $R_1$ do not have to comprise the same Pauli gates as long as Eqn.~\ref{eq:pauli-check-facts} is satisfied.

Recognizing that all errors can be decomposed into Pauli operators is a key concept in understanding PCS. By cleverly constructing Pauli sandwich checks dependent on $U$, anti-commutativity relations between the errors and checks can reveal errors in the ancilla qubits. Measuring 1 in any ancillas indicates that phase kickback occurred, meaning anti-commutation between the Paulis, and thus, a present error. As part of the PCS protocol, we discard this error-corrupted shot, leading to an increase in final distribution fidelity of our circuit of interest, $U$.

 Practical application of PCS requires consideration of tradeoffs. First, determining the Pauli unitary used in the check increases in complexity with the circuit. Second, Pauli check logic must be decomposed into the supported gate set of the QC, and as a result, large multi-qubit gates will end up being high-depth. 
 Third, the overhead of adding PCS into the circuit must not outweigh the corrective benefit in terms of gate error. Fourth, while increasing the number of check pairs improves error detection, it also introduces additional gate noise to the circuit and exponentially increases the required sampling size. We note that the exploration of an application's `check limit' is included in Ref.~\cite{langfitt2024pauli}. All of these limitations must be considered for practical implementation of PCS within quantum computing workflows.

\section{Ancilla-Free Pauli Checks}
\label{sec:pc-for-circ-opt}

Pauli checks are a useful tool for error detection in quanutum circuits. In addition to using Pauli checks for quantum error mitigation like in prior art, our proposed framework leverages Pauli checks in a novel manner to characterize quantum hardware during runtime for quantum program steering. We consider Pauli Check Sandwiching and Ancilla-Free Pauli Checks within this work.

We can estimate the postselection rate in PCS via an alternative method that does not require an ancilla and uses state preparation and measurements (SPAM) on the data qubits. We refer to this method as Ancilla-Free Pauli checks (AFPC). Note that we use AFPC only for the postselection information since in general the logical circuit is not preserved by AFPC in contrast to PCS. AFPC is a generalization of the one-sided checks introduced in Ref.~\cite{van2023singleside_check}. State preparation and measurements are used in tomography \cite{nielsen2010quantum} to extract channel information. AFPC can also be seen as a method that tracks a stabilizer state \cite{gottesman1998heisenbergrepresentationquantumcomputers} through the circuit. However, the correspondence between the PCS postselection rate and AFPC is our novel contribution.

Given that the noise channel is Pauli and the PCS method is ideal, i.e., the noise channel is only on the data qubits, the error syndrome from PCS only reveals information on the operations between the checks. Thus, state preparation and measurement gates lying outside the checks do not affect the postselection rate.
\begin{theorem}\label{thm:pcs_postselection_rate}
    For an ideal PCS scheme (noiseless checks and ancilla), the prepared input quantum state before the left checks and measurements after the right checks on the data qubits do not influence the postselection rate provided that the noise channel on the data qubits is Pauli.
\end{theorem}
An analytical proof of Theorem~\ref{thm:pcs_postselection_rate} is provided in the Appendix. Given Theorem~\ref{thm:pcs_postselection_rate}, the ancilla-free version of PCS can be derived from the stabilizer formalism \cite{gottesman1997stabilizercodesquantumerror}.

\begin{figure}
    \centering
    \includegraphics[width=0.9\linewidth]{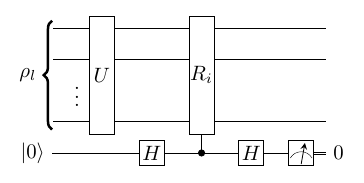}
    \caption{When $\rho_l$ is stabilized by $L_i$ the PCS circuit reduces to a Hadamard test circuit.}
    \label{fig:pcs_hadamard_meas}
\end{figure}

First consider the controlled Pauli operation $U_p=\op{0}\otimes I+\op{1}\otimes P.$
Let $\ket{\psi}$ be stabilized by $P$ and $\ket{+}=\frac{1}{\sqrt{2}}(\ket{0}+\ket{1})$. Then, the operation $U_p$ is a stabilizer of $\ket{+}\otimes \ket{\psi}$.
Thus, in AFPC we prepare the initial state to be a $+1$ eigenstate of $L_i$. Then, the PCS circuit reduces to a Hadamard test circuit as shown in Fig.~\ref{fig:pcs_hadamard_meas}.
The Hadamard test, checks if the state is in the $+1$ eigenspace of $R_i$. Since $R_i$ is a Pauli string, we can instead perform the test without an ancilla by measuring each of the data qubits in the $R_i$ Pauli basis and checking the parity. 
Thus, we have the following correspondence 
\begin{proposition}\label{prop:pcs_afpc_corres}
    For a Pauli noise channel on the data qubits, the post-selection rate of non-overlapping pairs (i.e., different target qubits) of ideal PCS checks can be extracted with ideal (noiseles SPAM) AFPC.
\end{proposition}
 Prop.~\ref{prop:pcs_afpc_corres} follows from Theorem~\ref{thm:pcs_postselection_rate} and the construction of AFPC. An example of the correspondence is provided in Fig.~\ref{fig:pcs-to-afpc}. Biasing the noise closer to Pauli can be achieved through twirling methods such as randomized compiling \cite{Wallman_2016noiseTailRadomCompiling}. Equivalence between the postselection rates of ideal PCS and ideal AFPC also happens when the prepared state in AFPC is equivalent to the input state in the PCS circuit. This happens, for example, when the left check is a tensor product of Pauli Z gates on the ground state.

\section{Methodology}
\label{sec:methods-for-distributions}

The previous sections describe how check data, or syndromes, from PCS or AFPC can inform whether or not an error occurred during the execution of a sandwiched circuit, $U$. Over many circuit samples, the amount of times that an ancilla is measured to be 1 can provide insight to the magnitude of quantum system noise. Consider the case where two identical circuits embedded with Pauli checks are executed in parallel on different regions of a QPU. If the check data is compared between two circuits, we can infer which execution contained more noise since a higher count of checks measuring as 1 implies that more error was present. In this Section, we first describe our techniques for embedding Pauli checks within quantum circuits. Next, we discuss how to leverage Pauli check data to guide quantum resource allocation for higher fidelity program outcomes. In the rest of this Section, we propose two key Pauli check protocols that estimate the noise profile of a target QC and guide output distributions to take most advantage of the best regions on a QPU.

\subsection{Evaluating Tradeoffs for Check Placement}

PCS can provide error mitigation benefits since post-selection can filter measured counts with errors. Both PCS and AFPC produce check data that can be used to learn about the noise profile of a QPU. We are interested in exploring the strengths of these approaches while being aware of the practical constraints of a realistic runtime operation environment. For this reason, we keep the PCS overhead minimal in terms of gate noise by only protecting single qubits that are on the edge of an application and thus assumed to be adjacent to an ancilla. This avoids 1) decomposing complicated unitaries for multi-qubit checks and 2) SWAP operations needed to transfer quantum state information to a non-adjacent ancilla. Because AFPC is lower overhead in terms of required gates and associated gate error, we are able to use them to check more qubits. 

We inject Pauli checks that use the same Pauli operation (i.e. $L_1,R_1 = X$ or $L_1,R_1 = Z$) to simplify the demonstration of Pauli checks injected into circuits. However, we note that this constraint can be relaxed as that Pauli checks do not need to be the same unitary for Pauli checks to be effective. For the benchmarks used in this work, it is probable that more optimal checks with higher circuit coverage exist, potentially providing more robust error mitigation and more significant fidelity improvements in certain noise models. However, we note that discovering optimal checks has poor scaling as $L_1$ and $R_1$ span more qubits. This serves as additional motivation to only use Pauli checks that check single qubits in both the ancilla and the ancilla-free case. For more complicated circuits (where Pauli checks cannot be found), sub-circuits instead of the full circuit can be examined instead.

\begin{figure*}[!th] \scriptsize \centering
	\includegraphics[width=\textwidth]{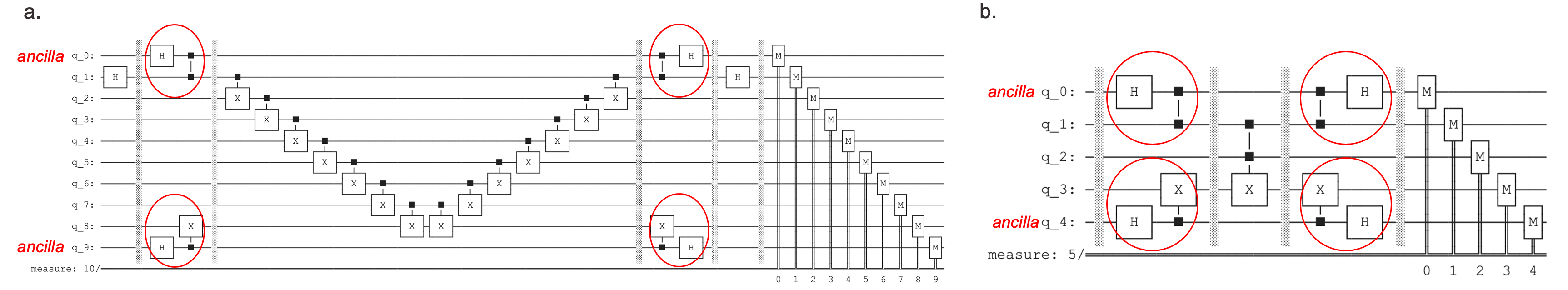}
	\caption{Benchmarks used for analysis of methods included (a) an eight-qubit GHZ / sensing circuit and (b) the three-qubit Toffoli operation. PCS checks are circled in red. The portion of each circuit protected by the PCS checks, $U$, lie in between the circled checks. Two additional ancilla qubits were needed for the two single-qubit PCS checks in both (a) and (b). AFPC versions of these circuits are also possible by inserting a single-qubit gate on the protected qubit that prepares the eigenstate of the Pauli operator in the PCS controlled-Pauli. An example of this conversion is included in Fig.~\ref{fig:pcs-to-afpc}.}
 \label{fig:benchmarks}
\end{figure*}

\subsection{Protocols for PC-enhanced Distributions}
\label{sec:protocols}

Many of today's QPUs feature more qubits than are required for target applications associated with general use. In this situation, numerous mapping permutations that assign logical to physical qubits are possible, and this observation introduces the questions of 1) how to make the best use of a quantum device and 2) how to maximize the exploration of the noise landscape to discover the regions that currently compute with highest fidelity. On one hand, one could prioritize using the `best' regions on-chip according to a mapping method guided by calibration data. This technique would serialized circuit execution to strategically target the historically low-error regions. However, this approach requires a priori knowledge of the properties of the QPU. In addition, serialized execution necessitates longer periods of dedicated hardware access. On the other hand, multi-programming to generate an ensemble distribution amplifies throughput via parallelization, potentially enabling lower-latency runtime. Additionally, it enables more of the device to be explored, potentially discovering unexpectedly high-performing regions, but it comes at the cost of also sampling less reliable qubits if the error profile varies drastically on the QPU. 

In our work, we assume that the number of qubits on chip is larger than the number of logical qubits required for an application. We also assume no prior knowledge of the noise profile on-chip in our main approach. We see that leveraging Pauli checks to learn quantum noise and boost program outcomes provides the substantial advantage of limiting dependence on data from the quantum hardware provider. While exploring the noise frontier of the QPU, we attempt to maximize hardware utilization as much as possible. 

\subsubsection{PCS Weighted Distribution}

We inject PCS into our target application, but only single-qubit checks on edge qubits are employed to minimize PCS gate count overhead. To maximally utilize hardware via multi-programming, a QC with $n$ physical qubits holds 
\begin{equation}\label{eq:n-threads}
    threads = \lfloor n/(q_{algorithm}+q_{ancilla}) \rfloor
\end{equation}

\noindent instances of a circuit running in parallel, assuming a graph with sufficient connectivity. In this equation, $q$ indicates qubits, both algorithm and ancilla, that are in the logical quantum circuit. Ancillas are only required in the PCS approach. We note while Eqn.~\ref{eq:n-threads} indicates a maximum number of parallel maps, any smaller set of qubit layouts may also be used with the techniques described in this Section.

Assuming the same shot budget in each local thread, values of the checks (i.e. PCS ancilla measurements of 1) are used during post processing to filter error-corrupted outcomes from each local thread's measurement distribution. This process creates a vector of error-mitigated counts for each thread, $c_i$. Circuits that encounter more noise will discard a greater count of shots. Discarded shot count for each local thread is represented with the scalar $r_{pcs,i}$. The percentage of discarded measurement outcomes for each thread, 

\begin{equation}
    d_{pcs,i} = \frac{r_{pcs,i}}{shots_{i}},
\end{equation}
\noindent gives insight into the underlying noise profile of the QC when different regions are compared. Each local $d_i$ is then used as a weight that scales $c_i$ to create $s_i$:

\begin{equation}\label{eq:scaled-counts}
    s_{pcs,i} = c_i * \frac{min(d_{pcs})}{d_{pcs,i}}.
\end{equation}

\noindent In Eqn.~\ref{eq:scaled-counts}, $min(d_{pcs})$ represents the minimum percentage of discarded shots from all the parallel threads. A cumulative distribution, $S_{pcs}$, is created from the sum of the $s_{pcs,i}$ vectors from the PCS-protected local threads, 

\begin{equation}\label{eq:cumulative-dist-pcs}
    S_{pcs}=\sum s_{pcs,i}.
\end{equation}

\subsubsection{AFPC Weighted Distribution}

AFPC can also be used to steer quantum program outcomes so that higher-performing QPU regions are favored while creating a final, ensembled distribution. The structure of the AFPC does not include an ancilla that is measured for syndrome information - syndrome information is read out directly from the qubit targeted by Pauli checks. This causes the AFPC approach for a weighted distribution to differ from that of the PCS approach in two ways. First, the upper bound of threads that can explore a QPU in parallel increases since $q_{ancilla} = 0$ in Eqn.~\ref{eq:n-threads}. Second, a payload circuit embedded with AFPC only returns syndrome data in contrast with a PCS circuit that produces an error mitigated program result along with syndrome data for a local thread. Because of this, the AFPC approach also needs data from a baseline run of the target circuit (i.e. a circuit without Pauli checks) in each of the local threads. To accurately capture noise data, the baseline and AFPC circuits must utilize the same physical qubits and should only differ because of any gates added for the AFPC. The syndrome data resulting from AFPC creates a weighted distribution using the baseline circuit outcomes, $b_i$, from local threads. The value used to scale local outcomes is determined by the count of AFPC measured outcomes where any syndrome was equal to one, $r_{af,i}$. A percentage of syndrome-generating outcomes can be calculated:

\begin{equation}
    d_{af,i} = \frac{r_{af,i}}{shots_{i}},
\end{equation}

\noindent which can then be used to create an AFPC weighted distribution,

\begin{equation}\label{eq:cumulative-dist-pcs}
    S_{af}=\sum b_i * \frac{min(d_{af})}{d_{af,i}}.
\end{equation}

\subsubsection{AFPC+threshold}
\label{sec:threshold}

The AFPC require two circuit evaluations: a baseline along with an AFPC injected equivalent. The check data can be used not only to assign weights to regions (i.e., regions with fewer detected errors receive higher weights in the combined distribution) but also to select which regions contribute to the combined distribution. This is done by ranking regions based on detected error rates and considering only the $n$ regions with the lowest detected errors. For $n=1$, the counts are taken exclusively from the single best region. For $n > 1$, Eq.~\ref{eq:cumulative-dist-pcs} is applied over the top $n$ regions. While the optimal number of regions to include cannot always be predicted beforehand, the experimental results in Section~\ref{sec:QAOA-results} of this paper indicate that taking the counts from only the top 1–3 regions typically yields the best performance for applications like QAOA. 

\subsubsection{Hybrid Pauli Check Approach} 
\label{sec:hybrid}
The primary benefit of the Pauli check techniques is that check data guides QC resource allocation rather than requiring backend-provided noise data for circuit mapping. However, the Pauli check techniques work synergistically with alternative quantum circuit optimizations that rely on noise data, if it is available. For example, \texttt{mapomatic} is a state-of-art tool that maps compiled quatum circuits to the best performing physical qubits of a QC according to device's calibration data~\cite{PRXQuantum.4.010327}. While a powerful optimizer, the quality of layouts that \texttt{mapomatic} returns are only as good as the data used. Drift in quantum systems causes actual machine parameters to stray from their calibrated metrics~\cite{dasgupta2021stability}, which will reduce the effectiveness of a \texttt{mapomatic} provided layout -- what was once the best may no longer be. However, Pauli checks can inform the quality of maps if multiple are returned from \texttt{mapomatic}. We propose in this approach, multiple \texttt{mapomatic} top picks could be generated and either the PCS or AFPC approaches could be applied to generate a PC +\texttt{mapomatic} weighted ensemble.

\section{Experimental Results}

In this Section, we apply the concepts from Sections~\ref{sec:pc-for-circ-opt} and~\ref{sec:methods-for-distributions} to evaluate the effectiveness of Pauli checks for error mitigation, quantum device characterization, and steering the quantum circuit distribution. This Section includes the three different experiments designed to evaluate PCS and AFPC enabled performance benefits on quantum circuits. First, the Pauli check methods described in Section~\ref{sec:methods-for-distributions} were evaluated in depolarizing noise channels. The second set of experiments targeted a QPU available in the IBM Quantum cloud in November 2024. The final experiment focused on the corrective benefits Pauli checks could provide to optimization problems. 

The benchmarks used during evaluation represented meaningful tasks in quantum computing:
\begin{itemize}
    \item \textbf{GHZ / Sensing} : The GHZ mirror circuit is used for evaluating quality of multi-qubit entanglement as well as in quantum sensing applications~\cite{wei2020verifying}.
    \item \textbf{Toffoli}: The Toffoli gate is a fundamental operator in quantum computing applications (and reversible logic), especially in its generalized form.
    \item \textbf{QAOA:} The Quantum Approximate Optimization Algorithm is a near-term quantum algorithm that finds approximate solutions to optimization problems.
\end{itemize}

\subsection{Pauli Checks in Depolarizing Noise}
\label{sec:eval-depolarizing}
We focus analysis on two circuits: GHZ and Toffoli. The GHZ circuit sizes include 2, 4, 6, and 8 qubits. The Toffoli gate has three qubits. Examples of the 8-qubit GHZ and 3-qubit Toffoli implementing PCS appear in Fig.~\ref{fig:benchmarks}. The translation from PCS to AFPC can be found in Fig.~\ref{fig:pcs-to-afpc}.

We are motivated to make best use of a QC with an unknown error profile at runtime though the application of Pauli checks to learn the noise landscape. Learned noise data then steer resource contributions to final circuit output distributions. To emulate this, we developed a noisy simulation implemented with Qiskit Aer~\cite{qiskit2024}. The test bench used during evaluation was an expansion of the model pictured in Fig.~\ref{fig:noise-profile} to emulate a QC with 60 regions where single-qubit depolarizing noise $p\in[0.0005,0.03]$ and two-qubit gates $2p$. One instance of each application was allocated per region so that 60 threads of each benchmark are executed in parallel. The circuits pictured in Fig.~\ref{fig:benchmarks} were transpiled to use the basis gate set that included $CX$, $X$, $SX$, and $RZ$.

\begin{figure}[t]
     \centering
         \includegraphics[width=0.99\linewidth,trim={0cm 0cm 0cm 0cm},clip]{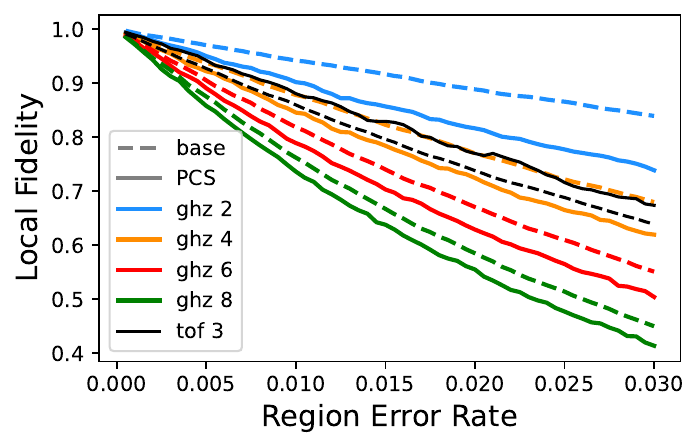}
        \caption{Local fidelity vs. region error rate of simulated QPU. Each of the sixty regions had a unique value of $p\in[0.0005,0.03]$ (two-qubit gates $2p$).}
        \label{fig:pcs-fidelities-v-error-local-simulated}
        
\end{figure}

\begin{figure}[t]
     \centering
         \includegraphics[width=0.99\linewidth,trim={0cm 0cm 0cm 0cm},clip]{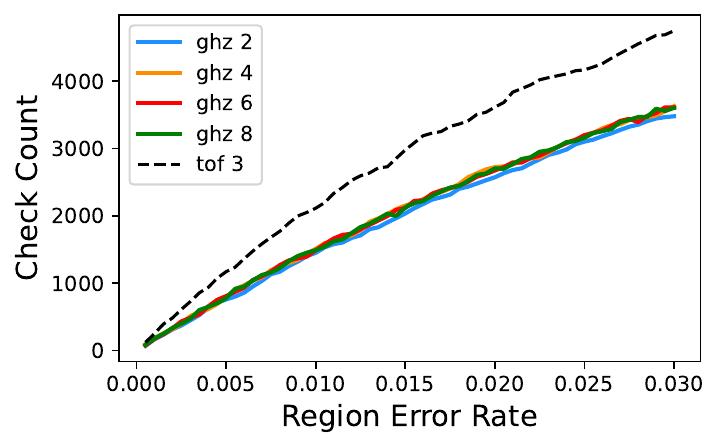}
        \caption{PCS check counts vs. QPU region error rate. 10,000 shots total were used in each experiment.}
        \label{fig:pcs-counts-simulated}
        
\end{figure}

\begin{figure}[t]
     \centering
         \includegraphics[width=0.99\linewidth,trim={0cm 0cm 0cm 0cm},clip]{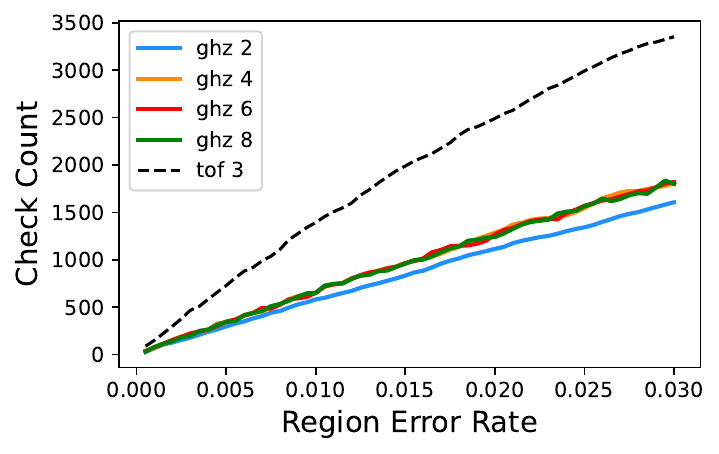}
        \caption{AFPC (edges only) check counts vs. QPU region error rate. 10,000 shots total were used in each experiment.}
        \label{fig:afpc-edge-counts-simulated}

\end{figure}

\begin{figure}[t]
     \centering
         \includegraphics[width=0.99\linewidth,trim={0cm 0cm 0cm 0cm},clip]{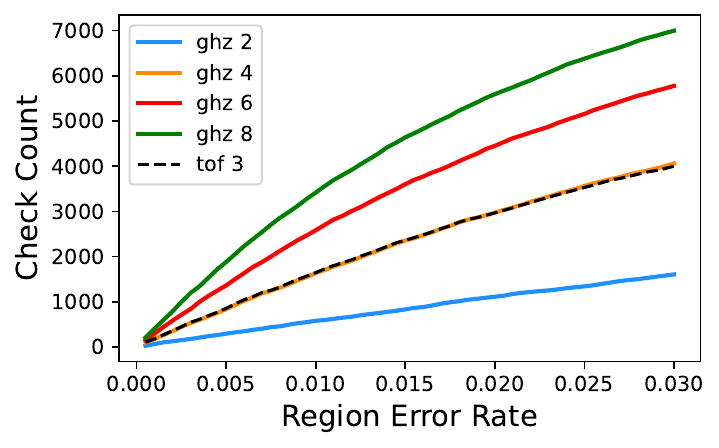}
        \caption{AFPC (all qubits) check counts vs. QPU region error rate. 10,000 shots total were used in each experiment.}
        \label{fig:afpc-counts-simulated}

\end{figure}

\begin{figure}[t]
     \centering
         \includegraphics[width=0.9\linewidth,trim={0cm 0cm 0cm 0cm},clip]{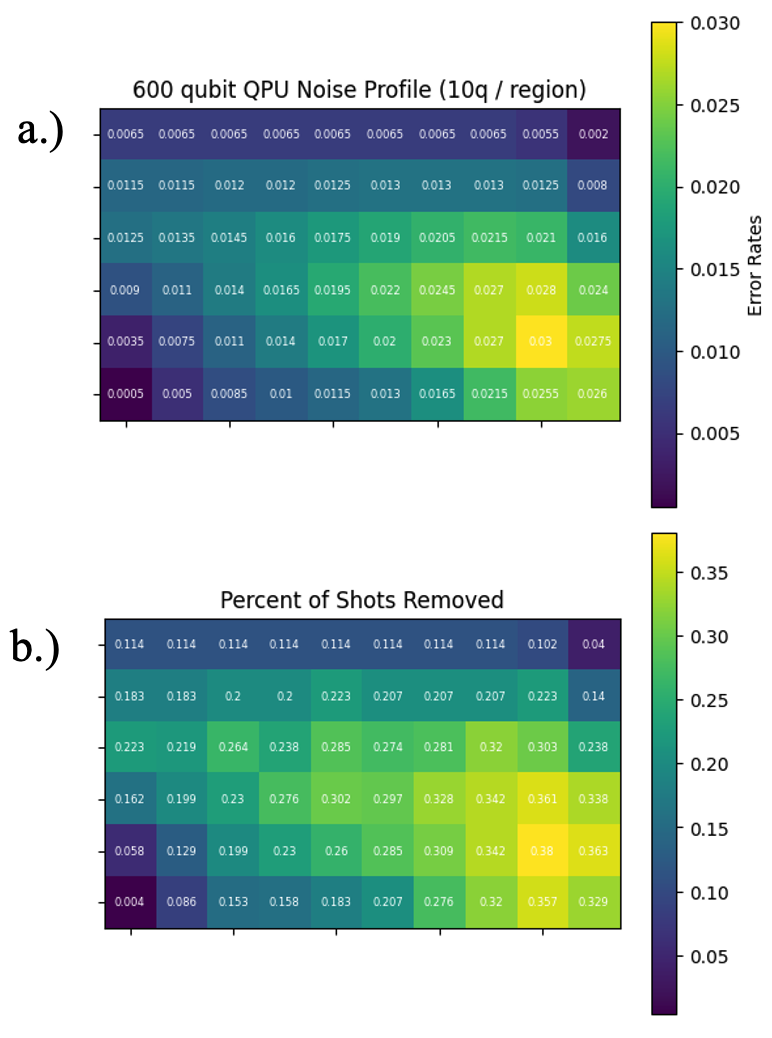}
        \caption{(a) Heatmap of an example QC chip noise profile, where each square is a 10 qubit region of the chip characterized by depolarizing noise, $p$. (b) Heatmap of plotting the percentage of shots removed after running 10,000 shots on PCS-protected 8-qubit GHZ circuit (seen in Fig.~\ref{fig:benchmarks}) on each region. The underlying QC chip is the same as pictured in (a).}
        \label{fig:random-noise-qc-and-heatmap}
        
\end{figure}

Relationships between fidelity and PC check / syndrome data were studied first. For each benchmark, four variations were generated: baseline with no Pauli checks, PCS on edge qubits, AFPC on edge qubits, and AFPC on all qubits. This experiment assumes parallel execution of all circuits of the same type, where one circuit is allocated to one of the 60 QPU regions. 10,000 shots are used for each experiment. 

First, we study error mitigated PCS vs. the baseline performance in all simulated regions. This is shown in Fig.~\ref{fig:pcs-fidelities-v-error-local-simulated} where dotted lines represent the no PCS baseline and solid lines represent PCS outcomes. We observe that while the Toffoli circuit reaps benefit from PCS within all regions, the GHZ outcomes suffer in all of the local (error mitigated) distributions. This is a clear demonstration that the tradeoffs of PCS must be carefully considered -- the error sensitivity of the circuit may not be able to withstand the effects of the additional 2q gate overhead. Despite PCS showing local fidelity degradation, we later apply the methods in Section~\ref{sec:methods-for-distributions} and find that parallelized PCS circuit outcomes and check data can be combined to produce significant fidelity improvements in the intelligently ensembled final distribution on the depolarizing error model.

Next, we study the relationships between Pauli check syndrome counts and device error rate for all Toffoli and GHZ circuits. Fig.~\ref{fig:pcs-counts-simulated} reports PCS data, Fig.~\ref{fig:afpc-edge-counts-simulated} reports AFPC placed only on the edge qubits (similar to PCS placement), and Fig.~\ref{fig:afpc-counts-simulated} reports AFPC when checks protect all qubits. These plots clearly show the relationships between syndrome data and region error. We highlight that Fig.~\ref{fig:afpc-counts-simulated} generally reports more checks counted for each circuit, which is an intuitive result as more of the circuit is protected by Pauli checks as compared to the circuits included in Figs.~\ref{fig:pcs-counts-simulated}and~\ref{fig:afpc-edge-counts-simulated}. The exceptions are the 3q Toffoli and 2q GHZ as their Pauli check protection remains minimally or completely unchanged going from edge to full AFPC.

To further instill confidence that Pauli checks can navigate a potentially unknown quantum compute landscape, a QPU was initialized using random error rates. Once again, a 600 qubit processor with 60 regions was targeted. A heat map representing this device is pictured in Fig.~\ref{fig:random-noise-qc-and-heatmap}(a). A 8-qubit GHZ circuit with PCS was executed in parallel for 10,000 shots, and the local percentage of shots removed in each region were used to generate a second heat map, Fig.~\ref{fig:random-noise-qc-and-heatmap}(b). The resemblance between the gradients of Figs.~\ref{fig:random-noise-qc-and-heatmap}(a) and~\ref{fig:random-noise-qc-and-heatmap}(b) shows that the syndrome data of the PCS checks can provide an estimate of the underlying QPU error rates. This gives us further indication that Pauli checks can be used in conjunction with multi-programming to scope out the noise profile of a QC, pushing towards higher fidelity results for quantum applications while minimizing dependence on characterization data.

\begin{figure}[t]
     \centering
         \includegraphics[width=0.99\linewidth,trim={0cm 0cm 0cm 0cm},clip]{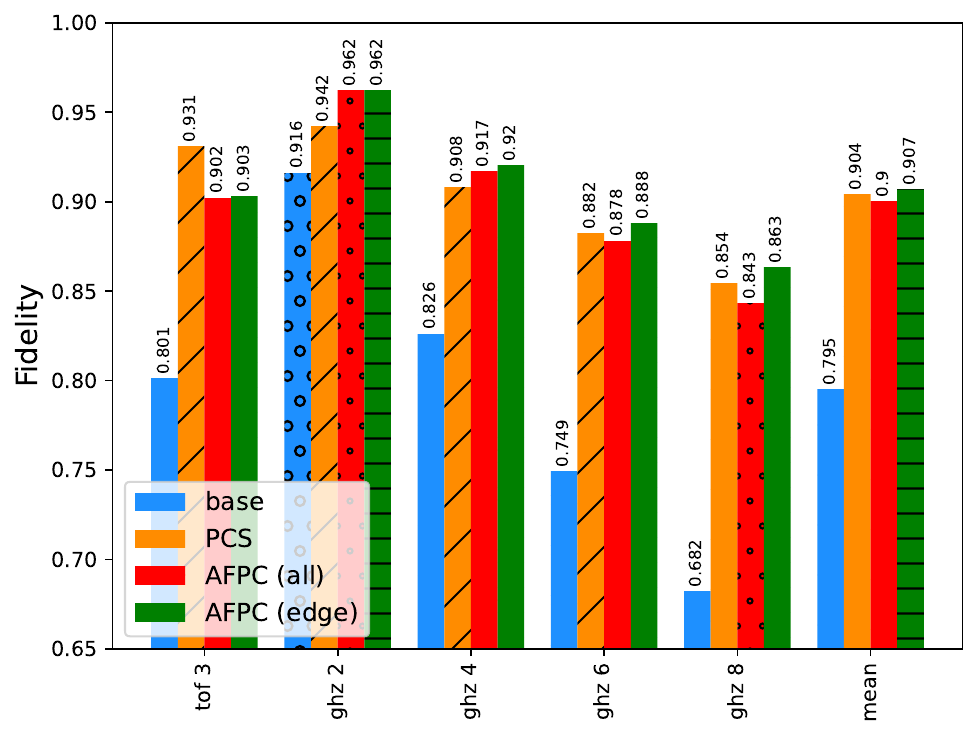}
        \caption{Fidelity results across different Pauli check methods.  }
        \label{fig:pcs-fidelities}
        
\end{figure}

 Fig.~\ref{fig:pcs-fidelities} shows the results for all circuits and Pauli check methods on the simulated QPU with a depolarizing noise model. The baseline comparison is generated from a basic ensemble of all QPU region outcomes. PCS and AFPC fidelities result from weighted distributions generated with the methods of Section~\ref{sec:methods-for-distributions}. The weighted ensembles created from both the PCS and the AFPC edge approaches demonstrate the best average improvement, $\sim14\%$.

\begin{figure}[t]
     \centering
         \includegraphics[width=0.95\linewidth,trim={0cm 0cm 0cm 0cm},clip]{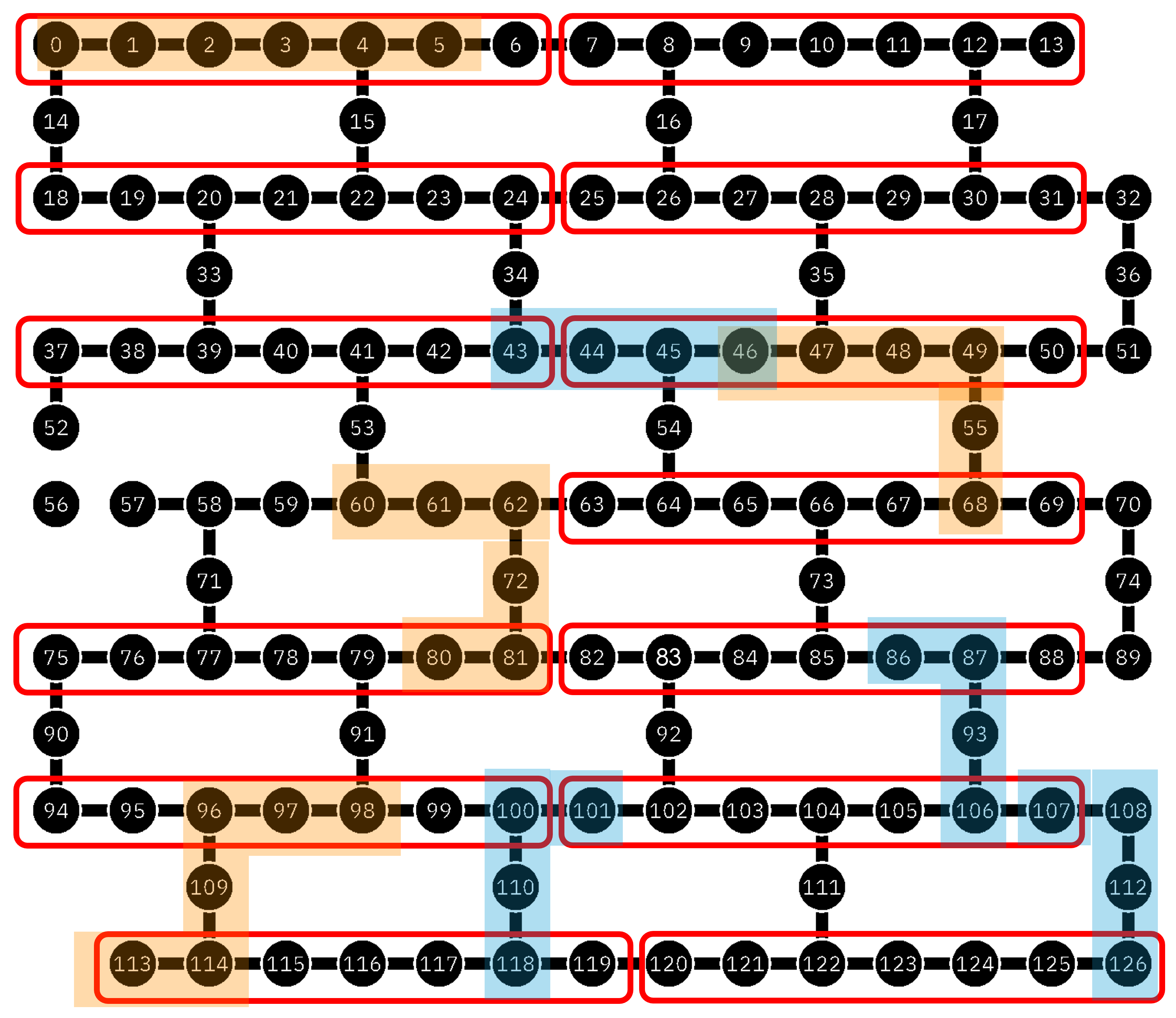}
        \caption{Layout of IBM Sherbrooke QPU. Target QPU regions for four-qubit GHZ experiments are emphasized. Regions circled in \textcolor{red}{red} (13 total) were used for logical to physcial qubit mapping for the noise-unaware parallelized GHZ. Regions highlighted with \textcolor{blue}{blue} are \texttt{mapomatic}-selected regions (top four) for the four-qubit GHZ. Regions highlighted with \textcolor{orange}{orange} are Mapomatic-selected regions (top four) for the four-qubit GHZ circuits with PCS. These regions were produced by Mapomatic on 14 November 2024.} 
        \label{fig:qpu-exp-circs-and-device}
        
\end{figure}

\subsection{IBM QPU Analysis}
\label{sec:IBM_QPU_Analysis}

\begin{figure*}[!th] \scriptsize \centering
	\includegraphics[width=\textwidth]{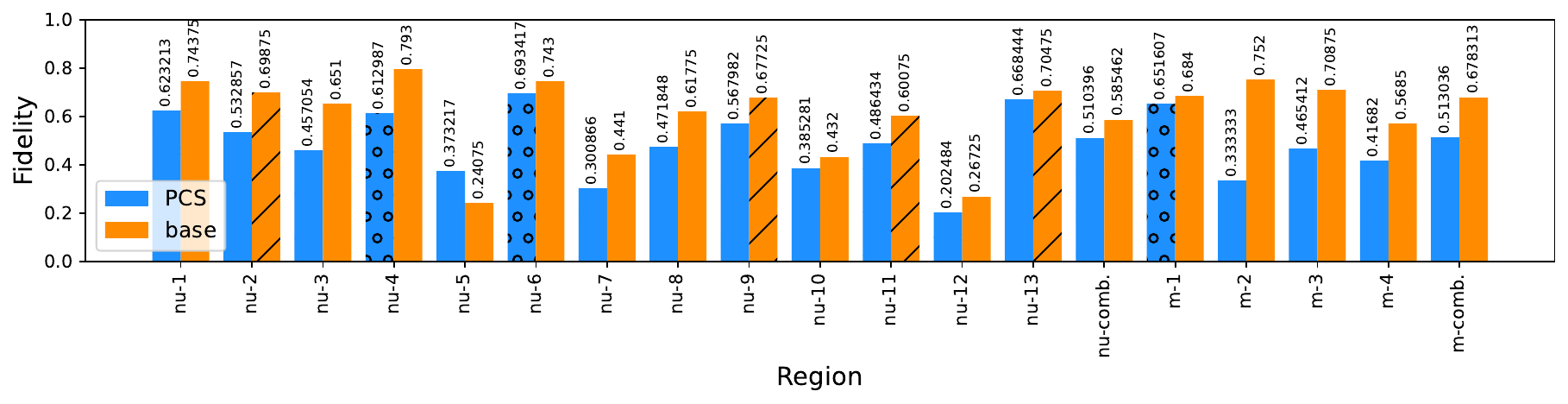}
	\caption{Four-qubit GHZ baseline and PCS results for the regions described in Fig.~\ref{fig:qpu-exp-circs-and-device}. The 13 noise-unaware regions are indicated by `nu' and the \texttt{mapomatic}-generated layouts are indicated by an `m'. The top \texttt{mapomatic} layout corresponds to `m-1'. Combined distributions are produced using the techniques outlined in Section~\ref{sec:protocols}.}
 \label{fig:QPU-circuits-4q}
\end{figure*}

\begin{figure*}[!th] \scriptsize \centering
	\includegraphics[width=\textwidth]{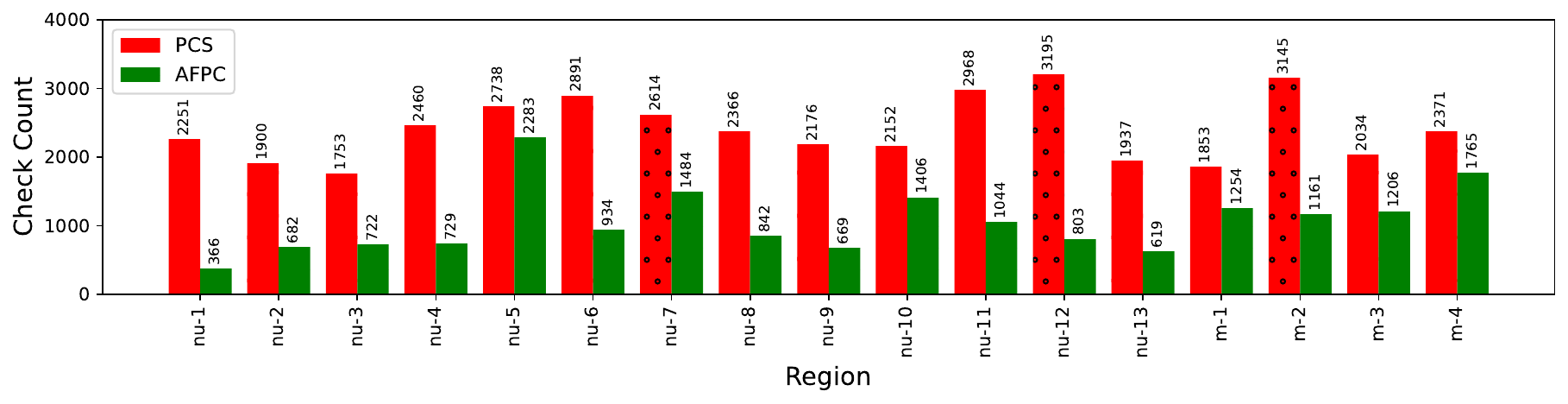}
	\caption{Four-qubit GHZ PCS and AFPC check counts for the regions described in Fig.~\ref{fig:qpu-exp-circs-and-device}. The 13 noise-unaware regions are indicated by `nu' and the \texttt{mapomatic}-generated layouts are indicated by an `m'. The top \texttt{mapomatic} layout corresponds to `m-1'. }
 \label{fig:QPU-circuits-4q-checks}
\end{figure*}

Pauli checks combined with a check-weighted distribution improved quantum circuit outcomes in a depolarizing noise channel. As a next form of evaluation, we tested PC and AFPC in a more irregular noise environment -- Pauli checks were placed in circuits run on real quantum hardware. Our real QPU experiments targeted \texttt{ibm\_sherbrooke} with the \texttt{ibm-q/open/main} instance that allows for 10 minutes of QC access per month. An image of this processor can be found in Fig.~\ref{fig:qpu-exp-circs-and-device}. 

We chose the GHZ/sensing circuit as the benchmark (Fig.~\ref{fig:benchmarks}(a)) for real QC evaluation as the circuit structure scales easily. The baseline was scaled from 2 - 4 qubits. Each circuit size had three variations: baseline with no Pauli checks, PCS on edge qubits (adds two extra qubits for the required ancilla), and AFPC on all qubits. All circuits of the same type were executed using 4,000 shots.

We were interested in how well Pauli checks could navigate a QPU noise landscape whenever arbitrary mappings were selected. To test this, the QC was divided into 13 uniform regions, indicated in red in Fig.~\ref{fig:qpu-exp-circs-and-device}, so that we could evaluate the quality of the Pauli checks weighted distribution when benchmarks were scheduled cross-chip. For consistency, we used the same regions for all circuit scales in our cross-QPU tests. We note that a region connecting physical qubits 56-62 was omitted due to a break in the device graph. In this test, we evaluated a naive ensemble, PCS, AFPC, and AFPC+threshold. In the threshold case (described in Section~\ref{sec:threshold}), the region with the lowest count of checks was sampled. We also compared our Pauli check approaches to two alternative error suppression approaches: noise aware mapping and noise aware ensemble. Noise aware mapping, referred to as `\texttt{mapomatic} top' chose the best region of circuits for the baseline using the best scoring layout returned by~\texttt{mapomatic}. Default options were selected during all \texttt{mapomatic} layout optimization procedures. Noise aware ensemble, referred to as `~\texttt{mapomatic} 4 ensemble' returned the top four maps from ~\texttt{mapomatic} and created an ensembled distribution from those circuit outcomes. Four maps were chosen in this approach as it was the default setting in Ref.~\cite{tannu2019ensemble}. Finally, we implemented a hybrid, \texttt{mapomatic}-guided Pauli check approach where all the Pauli check methods previously mentioned were used only in the four regions selected by \texttt{mapomatic}. The Pauli check hybrid method is described in Section~\ref{sec:hybrid}.

First, we study the results from the four-qubit GHZ circuit. We examine the performance of the 13 regions of the noise unaware mapping to gain a better understanding of performance cross-chip. We also analyze the outcomes guided by \texttt{mapomatic}. The regions selected by \texttt{mapomatic} for the four qubit GHZ are found in Fig.~\ref{fig:qpu-exp-circs-and-device} -- the baseline and the PCS injected circuit are emphasized in blue and orange, respectively. The baseline and PCS fidelity results for each region are found in Fig.~\ref{fig:QPU-circuits-4q}. These results show that the performance varies significantly cross-chip. When focused on the baseline, we observe that the top \texttt{mapomatic} pick does not have the higest fidelity - seven other mappings demonstrate better performance. Comparing the baseline to PCS, we see that the error overhead associated with the four additional two-qubit gates required for the Pauli checks outweigh the corrective potential of PCS error mitigation. This is consistent with the results we saw in the depolarizing channel for GHZ (Fig.~\ref{fig:pcs-fidelities-v-error-local-simulated}). Fig.~\ref{fig:QPU-circuits-4q-checks} shows the check counts for the PCS and AFPC circuits mapped to the 13 regions of Fig.~\ref{fig:qpu-exp-circs-and-device} circled in red. We see in every case that check counts are lower for full AFPC as compared to PCS, which is opposite from our findings in Section~\ref{sec:eval-depolarizing}.  

We find final fidelity results for each benchmark of each type in Table~\ref{tab:qpu-results}. In this table, the highest fidelity approach for each circuit size is highlighted in orange while the second highest fidelity is highlighted in blue. Although some cases are within noise margins, technique on Pauli checks resulted in the highest fidelity in each case. We find that when the AFPC utilizes a threshold, it has a high probability of finding a quality mapping from a selection of provided maps. We note that as the circuit size gets larger, the Pauli check guided techniques are able to distance themselves from the baseline.

\begin{figure}[t]
     \centering
         \includegraphics[width=0.99\linewidth,trim={0cm 0cm 0cm 0cm},clip]{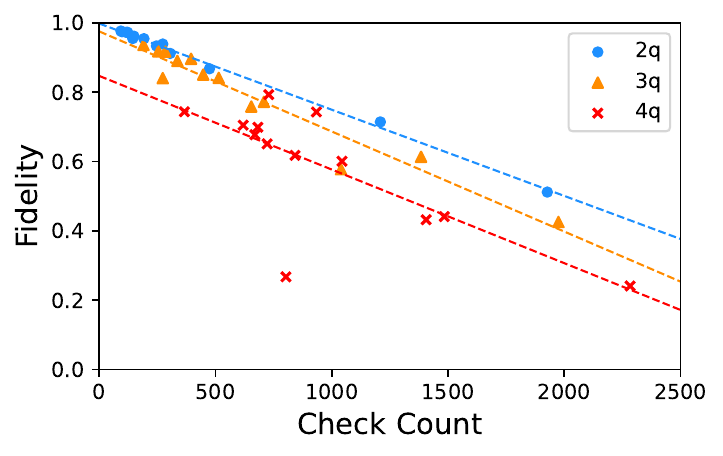}
        \caption{QPU results for fidelity vs. AFPC check counts for 2-4 qubit GHZ circuits. Here, a clear relationship is seen between region performance and syndrome data from the AFPC circuits.}
        \label{fig:qpu-exp-fid-vs-counts}
        
\end{figure}

As a final analysis, we examine how well Pauli checks learn the error profile of a real quantum machine. We do this by comparing the AFPC check count with baseline circuit fidelity for the 13 arbitrary regions that span IBM Sherbrooke. This analysis is included in Fig.~\ref{fig:qpu-exp-fid-vs-counts}. Here we see a direct correlation between baseline fidelity and AFPC check count for all GHZ benchmarks. This result agrees with the relationships found in the depolarizing noise model study (Fig.~\ref{fig:afpc-counts-simulated}).

\begin{table*}[t]
  \centering
  \begin{tabular}{c|c|c|c|c|c|c|c|c|c|c}
    Qubits & \multicolumn{4}{c|}{13 Regions (noise-unaware)}  & \multicolumn{4}{c|}{Mapomatic-guided PCs}  & Mapo. Top & Mapo. 4 Ens.  \\
    \hline
    \cellcolor{black} & Base Ens. & \textbf{PCS} & \textbf{AFPC} & \textbf{AFPC+Thr.}  & Base Ens. & \textbf{PCS} & \textbf{AFPC} & \textbf{AFPC+Thr.} & \cellcolor{black}  & \cellcolor{black}   \\
    \hline
    2& 0.8958&	0.9042&	0.9532&	0.9765&	0.9767&	0.9607&\cellcolor{cyan} 0.9771&\cellcolor{orange}0.9783&\cellcolor{orange}0.9783&	0.9767 \\
    3& 0.7876&	0.7575&	0.8580&\cellcolor{orange}0.9365&	0.9219&	0.7035&\cellcolor{cyan}	0.9236&	0.9218&	0.9133&	0.9219 \\
    4& 0.5855&	0.5104&	0.6328&\cellcolor{cyan}	0.7438&	0.6783&	0.5130&	0.6884&\cellcolor{orange}0.7520&	0.6840&	0.6783 \\

  \end{tabular}
  \caption{Fidelity results of GHZ benchmark on IBM QC Sherbrooke. PC techniques (\textbf{bold}) are compared to the baselines of ensemble and Mapomatic approaches. At each circuit size, highest fidelity is indicated in \textcolor{orange}{orange} and second highest fidelity is indicated in \textcolor{blue}{blue}. We find that when the AFPC utilizes a threshold, it has a high probability of finding a quality mapping from a selection of provided maps.}
  \label{tab:qpu-results}
\end{table*}

\subsection{Pauli Checks for Quantum Approximate Optimization}
\label{sec:QAOA-results}

\begin{figure}[h]
    \centering
    \includegraphics[width=\linewidth]{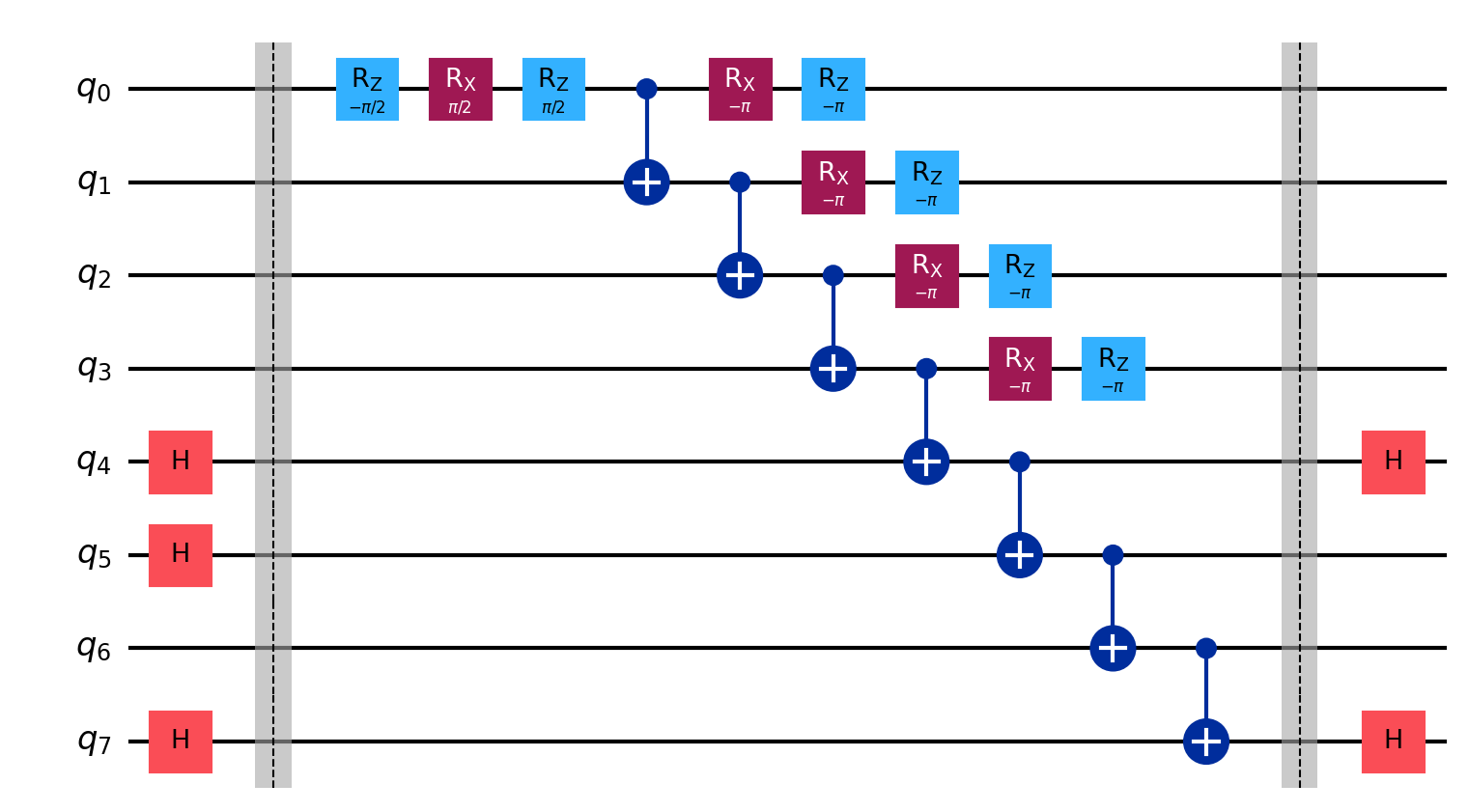}
    \caption{Quantum circuit diagram of the hardware-efficient QAOA ansatz with AFPC. The checks are indicated by the Hadamard gates positioned outside the barriers.}
    \label{fig:qaoa_circuit}
\end{figure}

In this Section, we present results for a realistic application circuit executed on the FakeBrisbane mock backend provided by Qiskit. We focus on the hardware-efficient Quantum Approximate Optimization Algorithm (QAOA) ansatz, as described in \cite{Moll_2018}, and use pre-optimized parameters that were obtained under a noiseless setting. The circuits evaluated are of size 8 and 10 qubits with depths ($p$) of 1 and 2. Two checks are applied to the QAOA circuit: one targeting the controlled-X gate on the bottom-edge qubit and another targeting a controlled-X gate on one of the center qubits. The circuit is then converted to its AFPC variant, which is used to calibrate the error rates of the candidate regions. An example of the pre-optimized $8$-qubit QAOA circuit with AFPC is shown in Fig.~ \ref{fig:qaoa_circuit}. It is important to note that better performance is expected with more checks. However, these experiments demonstrate that even a small number of checks can provide significant performance boosts, even for larger systems. This is particularly valuable because identifying suitable checks becomes increasingly challenging as system size grows \cite{gonzales2023quantum}.

We compare two methods for evaluating performance improvement over the baseline: AFPC and AFPC + threshold. In the AFPC + threshold approach, we evaluate the top 1, 2, and 3 regions, where the counts for the top 2 and top 3 regions are combined using the weighted distribution in Eq.~\ref{eq:cumulative-dist-pcs}. The PCS ensemble approach is excluded from this comparison, as it consistently underperformed relative to the baseline under the assumed noise model. This was likely due to the noise overhead introduced by the SWAP operations when mapping the checks from PCS to the physical device qubits if algorithm qubits protected by a check could not be placed next to an ancilla qubit.

\begin{figure}[h]
    \centering
    \includegraphics[width=\linewidth]{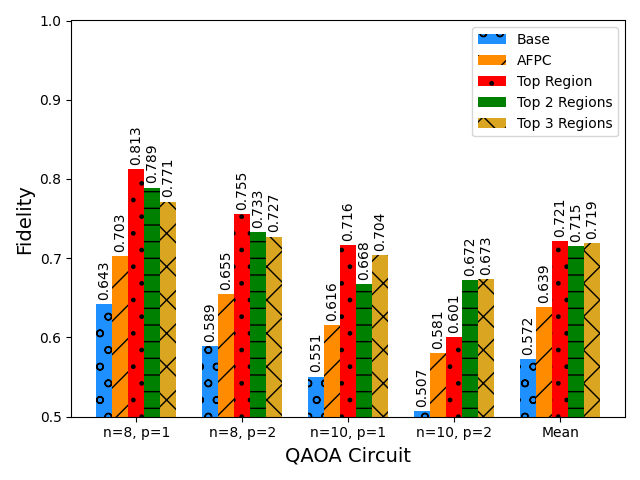}
    \caption{Fidelities for each AFPC method across various QAOA circuits. `Mean' represents the average fidelities across all four circuit instances.}
    \label{fig:real_app_results}
\end{figure}

Fig. ~\ref{fig:real_app_results} presents the results for the various QAOA circuits. The initial mappings we used were selected in a similar manner as for the IBM Sherbrooke backend detailed in the previous subsection. Across all test circuits, we observe substantial improvements in fidelities when comparing the top region(s) selected by AFPC to the baseline circuit ensemble. For the 8-qubit circuits, the average fidelity improvement is approximately $27\%$, while for the 10-qubit circuits, fidelity improves by approximately $33\%$. Also, we can see that the best performance is sometimes achieved using only the top region, whereas in other cases, combining the top 2 or 3 regions results in greater improvements (see the fidelities for QAOA with $n=10$ in Figure~\ref{fig:real_app_results}).

\section{Discussion and Future Directions}

Our experiments found that while PCS ensembling could result in performance improvements, tradeoffs with gate error needed to be carefully considered. In noisy environments, like IBM Sherbrooke or the FakeBrisbane mock backend, PCS was not helpful. On the other hand, AFPC was shown to have more uniform performance. An important contribution of this work is highlighting the ability of AFPC to coarsely calibrate error rates across QPU regions with minimal resource overhead. While identifying suitable Pauli checks can become challenging for larger circuits, the AFPC method introduces only single-qubit gates, which add relatively little error to the circuit. For very complicated circuits (where Pauli checks cannot be found), sub-circuits instead of the full circuit can be examined to still learn noise that matters to a quantum application. Despite this low overhead, AFPC was able to accurately identify the best-performing regions, resulting in substantial fidelity improvements for circuits as large as 10 qubits, requiring only two single-qubit checks. This efficiency makes AFPC a practical and scalable solution for error mitigation and hardware optimization in near-term devices.

Another key advantage of Pauli checks is their flexibility in quantum circuit mapping. As demonstrated in Section~\ref{sec:IBM_QPU_Analysis}, Pauli checks, when leveraged in a hybrid approach, can be combined with prior knowledge or noise distribution estimates to enhance performance. For example, we used \texttt{mapomatic} to identify approximate optimal regions, which were further refined with Pauli checks to achieve significant improvements in fidelity. This integration highlights the potential for combining hardware-level noise characterization tools with PCS to optimize qubit mappings. An area for future work would be to explore other quantum circuit optimization techniques that work synergistically with Pauli checks.

A potential method for future exploration is a dynamic shot allocation approach. In this method, given a shot budget, shots are incrementally sent to each region of the chip. For each region, errors are monitored in real time (by either the PCS or AFPC circuit), and once the the number of discarded shots exceeds a specified threshold, no additional shots are allocated to that region. The remaining shot budget is then gradually focused towards regions with lower error rates, allowing resources to be dynamically steered toward the best-performing regions on chip. This approach contrasts with the Pauli check methods demonstrated in this work, where a fixed number of shots are allocated per region to estimate the least noisy regions. While dynamic shot allocation offers the potential for more efficient resource usage, its effectiveness depends on accurately selecting the discard threshold, as this parameter influences the likelihood of identifying the best regions. We anticipate that the optimal threshold will vary across devices and applications, and future research should focus on identifying domain-specific thresholds to ensure consistent and reliable results.

\section{Conclusion }

This work highlights the versatility and potential of Pauli checks in improving quantum program performance, particularly in the resource-constrained environments of current quantum devices. By combining Pauli checks with multi-programming, we present a framework that not only mitigates errors in individual circuit distributions but also enables intelligent resource allocation and region selection across QPUs.

Beyond their role in error mitigation, Pauli checks can guide efficient qubit mapping and optimize hardware utilization, even in scenarios with little to no prior knowledge of the noise distribution. This capability for "noise-unaware" mapping is especially valuable for real-time hardware characterization or when operating under a strict shot budget. Looking forward, expanding these methods to include dynamic resource allocation strategies, such as the proposed dynamic shot allocation, could further improve resource efficiency and performance. Additionally, extending these approaches to more complex quantum algorithms beyond QAOA and scaling them to larger systems represents an exciting avenue for future research, paving the way for scalable, efficient, and easily adoptable quantum computing solutions.

\section{Appendix}
Proof of Theorem~\ref{thm:pcs_postselection_rate}.
\begin{proof}
    We denote the arbitrary input density matrix as $\rho$. First, consider a single pair of ideal checks sandwiching a noisy circuit $U$. Let the noisy channel be $\mathcal{E}(\psi)=\sum_iE_i\psi E_i^\dagger.$ The postselection probability is \cite{gonzales2023quantum} 
    \begin{align}\label{eq:postselectPCS}
        p=\frac{1}{4}\sum_i\text{tr}([RE_iR+E_i]U\rho U^\dagger[RE_i^\dagger R+E_i^\dagger]).
    \end{align}
     Let $E_i'=\frac{RE_iR+E_i}{2}$. Then Eq.~\eqref{eq:postselectPCS} can be written as \begin{align}\label{eq:postselectPCS2}
         p=\text{tr}\left(U\rho U^\dagger\sum_iE_i'^\dagger E_i'\right).
     \end{align} 
     Expand $E_i$ in the $+1$ Pauli basis as $E_i=\sum_ja_{ij}P_j$. Notice that the Paulis that anticommute with $R$ vanish. Let the set of +1 Paulis that commute with $R$ be indexed by the set S. Thus $E_i'=\sum_{j\in S}a_{ij}P_j$. From direct calculation
    \begin{align}
        &\sum_iE_i'^\dagger E_i=\sum_{ij\in S,k\in S}a^*_{ij}P_ja_{ik}P_k\\
        &=\sum_{i}\left(\sum_{j\neq k|j\in S,k\in S}a^*_{ij}a_{ik}P_jP_k+\sum_{j=k\in S}|a_{ij}|^2I\right).
    \end{align}
    Finally, if $\sum_{j\neq k|j\in S,k\in S}a^*_{ij}a_{ik}P_jP_k=0$, we have that $\sum_iE_i'^\dagger E_i=\sum_{j\in S}|a_{ij}|^2I$. From substitution into Eq.~\eqref{eq:postselectPCS2}, we have that $p=\sum_{j\in S}|a_{ij}|^2$ $\forall \rho.$ Pauli channels have the property that \newline $\sum_{j\neq k|j\in S,k\in S}a^*_{ij}a_{ik}P_jP_k=0$ since $E_i=a_{i}P_i$. The extension to multiple checks follows.
\end{proof}

\section{Acknowledgments}
JL, AG, and ZHS acknowledge support by the Q-NEXT Center. NH was partially supported by NSF CCF-2119069. The submitted manuscript has been created by UChicago Argonne, LLC, Operator of 
Argonne National Laboratory (``Argonne''). Argonne, a U.S.\ Department of 
Energy Office of Science laboratory, is operated under Contract No.\ 
DE-AC02-06CH11357. 
The U.S.\ Government retains for itself, and others acting on its behalf, a 
paid-up nonexclusive, irrevocable worldwide license in said article to 
reproduce, prepare derivative works, distribute copies to the public, and 
perform publicly and display publicly, by or on behalf of the Government.  The 
Department of Energy will provide public access to these results of federally 
sponsored research in accordance with the DOE Public Access Plan. 
http://energy.gov/downloads/doe-public-access-plan.

\bibliographystyle{plain}
\bibliography{ref}

\end{document}